\documentclass[a4paper]{spie}  

\usepackage{amsmath,bm}
\usepackage{graphicx}
\usepackage{url}

\usepackage{multirow}
\usepackage{xcolor}
\usepackage{lscape}
\usepackage{textcomp}
\usepackage{txfonts}
\usepackage{color}
\usepackage{mathtools}
\usepackage[colorlinks=true, allcolors=blue]{hyperref}
\usepackage{sidecap}
\usepackage{subcaption}

\title{ngVLTI: Cost and Feasibility of a Four-Telescope VLTI Expansion for Milliarcsecond-Scale Imaging}
\author[1]{Sylvestre Lacour}
\author[2]{Johan Kosmalski}
\author[2]{Michael Mueller}
\author[3]{Célia Desgrange}
\author[4]{Oscar Carrión-González}
\author[1]{Mathias Nowak}
\author[1]{Elsa Huby}
\author[2]{Julien Woillez}
\author[2]{Jens Kammerer}
\author[5]{Jean-Baptiste Le Bouquin}
\author[2]{Jason Spyromilio}
\affil[1]{Laboratoire d'Instrumentation et de Recherche en Astrophysique, Observatoire de Paris, PSL University, CNRS, Sorbonne Universit\'e, Universit\'e Paris Cit\'e, 5 place Jules Janssen, 92195 Meudon, France}
\affil[2]{European Southern Observatory, Karl-Schwarzschild-Stra\ss{}e 2, 85748 Garching bei M\"unchen, Germany}
\affil[3]{European Southern Observatory, Alonso de C\'ordova 3107, Vitacura, Regi\'on Metropolitana, Chile}
\affil[4]{Max-Planck-Institut f\"ur Astronomie, K\"onigstuhl 17, 69117 Heidelberg, Germany}
\affil[5]{Universit\'e Grenoble Alpes, CNRS, IPAG, 38000 Grenoble, France}
\begin{document}
\maketitle

\begin{abstract}
We assess the cost and technical feasibility of extending the Very Large Telescope Interferometer (VLTI) through the addition of four 8\,m-class Unit Telescopes (UTs), an upgrade we refer to as the new generation VLTI (ngVLTI). Such an upgrade would provide a dense and homogeneous \textit{uv} coverage with baselines up to 220\,m, enabling true imaging at milliarcsecond angular resolution across science cases ranging from Solar System bodies to distant active galactic nuclei. Because each image is reconstructed at a single wavelength, repeating the reconstruction across the spectral channels of the instrument would deliver spectral-imaging cubes -- a qualitatively new capability for the VLTI, bringing it close to the imaging power of ALMA at near-infrared wavelengths. Motivated by this scientific case, we examine a compact telescope concept consisting of a fast ($f/0.64$), segmented, parabolic primary mirror feeding a subterranean coud\'e focus compatible with the existing VLTI infrastructure. We summarise the optical design, which achieves diffraction-limited performance over a 1\,arcmin field of view and a well-matched reimaged pupil, discuss the mechanical trade-offs behind the choice of a 60-segment, 1.2\,m primary -- a mirror mass of $\approx$13.5\,tons and an altitude moving mass of $\approx$50\,tons -- and quantify the gravitational flexure of the telescope structure and its resulting optical sensitivity as a function of pointing elevation. We then present a back-of-the-envelope cost estimate of order 80\,M\texteuro\ per telescope (2026 prices), broken down into the segmented primary, adaptive secondary, coud\'e train, mount, enclosure, and ancillary instrumentation. Benchmarked against the historical costs of Keck, Gemini, and the VLT UTs, and scaled to a four-telescope array, the proposed upgrade appears both cost-competitive and technically achievable, offering a long-term perspective for Paranal Observatory in the ELT era.
\end{abstract}


\section{INTRODUCTION}
\label{sec:intro}

The Plateau de Bure Observatory began as a three-antenna radio interferometer \cite{1992A&A...262..624G}. With its transformation into NOEMA (NOrthern Extended Millimeter Array) \cite{Chenu2016}, it reached its full scientific potential following the inauguration of its seventh antenna in September 2014. The transition from an array primarily suited to model fitting to a true imaging instrument is, of course, not abrupt. Nevertheless, experience shows that the regime of seven to eight elements represents a critical threshold: prior to NOEMA, Plateau de Bure largely served as a technical and scientific pathfinder for ALMA; afterwards, it became a highly productive imaging observatory in its own right.

We argue that the VLTI is now at a similar turning point. With the advent of GRAVITY \cite{gravity2017}, the VLTI has reached technological maturity and demonstrated its unique scientific power. However, to fully deliver on its potential for the community, the VLTI requires additional telescopes. Robust imaging capabilities are essential to move beyond highly parameterised models and to access physical regimes that cannot be constrained with sparse interferometric data alone. Crucially, interferometric images are reconstructed one wavelength at a time; a dense array would therefore not only deliver monochromatic images, but -- by repeating the reconstruction over the many spectral channels of modern instruments -- allow the VLTI to routinely produce spectral-imaging cubes. This would be an entirely new capability for the VLTI, placing it almost on par with ALMA in its ability to map morphology and kinematics jointly, at near-infrared wavelengths and milliarcsecond resolution.

The scientific case for such an upgrade -- summarised in Sec.~\ref{sec:science} -- is broad and compelling, and has been discussed extensively elsewhere\footnote{See \url{https://horizons-olbin.sciencesconf.org/}.}. The central question addressed by this paper is therefore a different one: \emph{how much would it cost, and how feasible is it, to upgrade the VLTI with four additional Unit Telescopes?} We describe a specific, compact telescope concept (Sec.~\ref{sec:optics}) capable of meeting this goal within the existing Paranal infrastructure, and we derive an itemised, order-of-magnitude cost estimate for it (Sec.~\ref{sec:cost}), explicitly stating the assumptions on which this estimate rests.

\section{SCIENTIFIC MOTIVATION}
\label{sec:science}

\begin{figure}[htp]
   \begin{center}
      \includegraphics[width=0.86\linewidth]{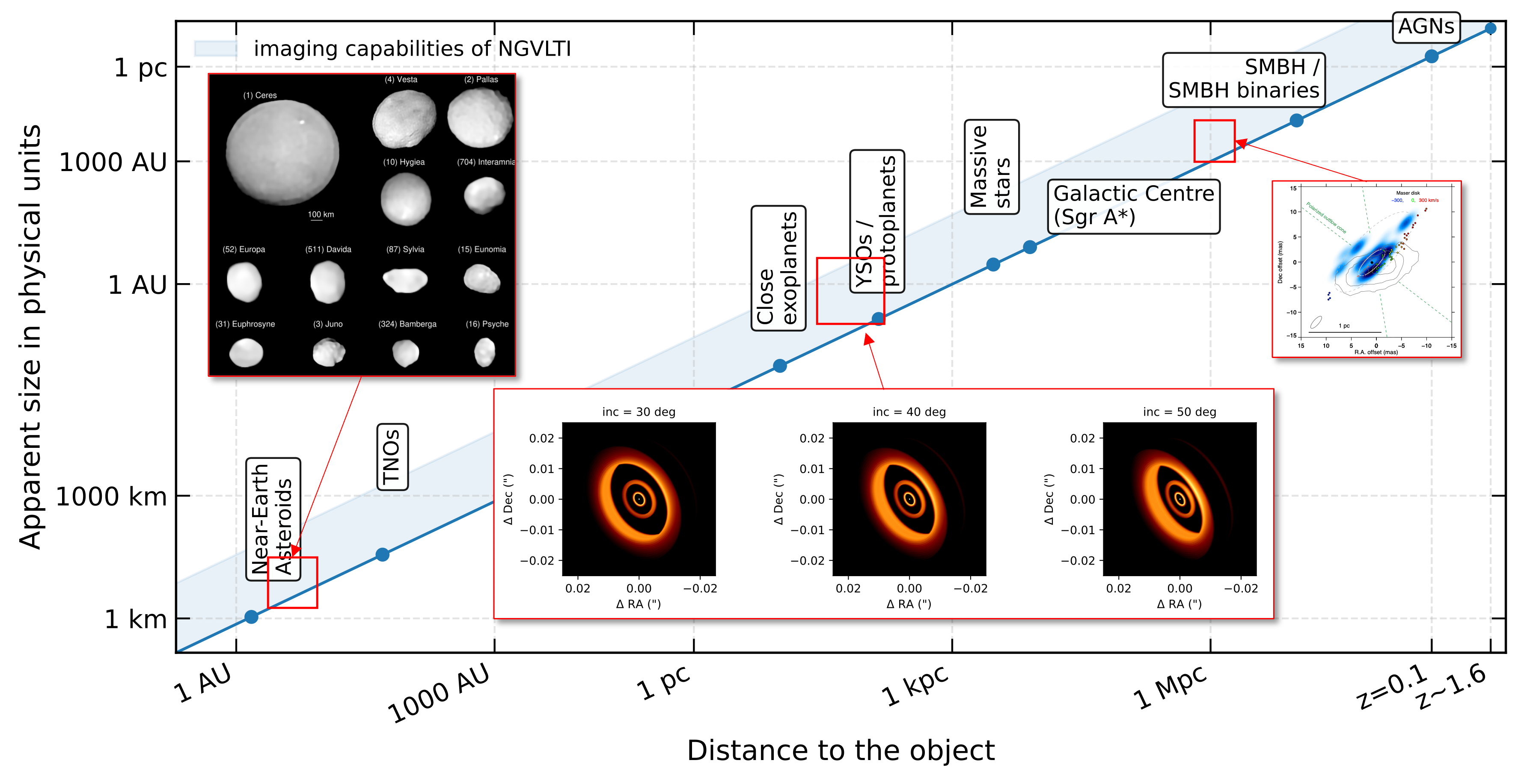}
\caption{Physical spatial resolution of ngVLTI as a function of distance.  Boxes indicate the approximate distance regimes relevant for key science cases, spanning Solar System objects (near-Earth asteroids and trans-Neptunian objects), nearby and Galactic targets (exoplanets, young stellar objects, massive stars, and the Galactic Centre), and extragalactic sources (supermassive black holes, binary SMBHs, and active galactic nuclei). Inserted are 3 science cases: small bodies \cite{2021A&A...654A..56V}, AGN imaging\cite{2020A&A...634A...1G}, and an illustration of a protoplanetary disk. These images are based on radiative-transfer simulations performed with the \textsc{MCFOST} code \cite{2006A&A...459..797P,2009A&A...498..967P}. The disk exhibits multiple concentric rings located at radial distances of approximately 0.3--0.5\,au, 0.8--1.0\,au, and 2--3\,au, representative of sub-au to few-au scale structures associated with planet formation processes. Such spatial scales are directly accessible with milliarcsecond-resolution interferometric imaging, highlighting the potential to image planet--disk interactions in nearby star-forming regions.
}  \label{fig:science}
   \end{center}
\end{figure}

Proposals to expand the VLTI beyond its current number of telescopes are not new. Recent white papers submitted to ESO's Expanding Horizons call also justify building more telescopes: for time-domain imaging of young stellar objects, relying on the existing 1.8\,m Auxiliary Telescopes restricts the accessible sample to a small fraction of the brightest targets, motivating proposals for six or more new, larger telescopes dedicated to reactive, time-domain interferometry \cite{2026arXiv260203401S}. For exoplanets, detectability would benefit from baselines longer than those currently available at Paranal, and a UT5 positioned to the south, increasing the baseline lengths and filling the existing gap in north-west-oriented baselines \cite{2025A&A...694A.277L}. Other projects have also been proposed to reach kilometer-scale baselines, which would allow high-resolution parametric fitting and $0.1\,\mu$as astrometry \cite{2024sf2a.conf..183B}.

But a key strength of the proposed facility lies in the diversity of scientific cases it can address, as well as in the wide range of physical scales it can probe, from Solar System bodies to objects at extreme redshifts.
Probing this wide range of spatial scales requires high sensitivity, which is a key driver of the proposed concept. Surface brightness sensitivity is particularly critical, from faint Solar System objects such as trans-Neptunian objects to distant active galactic nuclei. Observing these low surface brightness objects demands a large collective collecting area, which can only be provided by 8\,m-class telescopes.
In this context, the development of the ELT programme offers a unique opportunity: the use of primary mirrors of similar size and a common segmented-mirror technology would allow significant technological synergy, reducing development risks and overall costs through shared design, manufacturing, coating, and operational experience -- a connection we quantify further in Sec.~\ref{sec:cost}.
Beyond its scientific impact, this project would provide a long-term perspective for the Paranal Observatory, ensuring its continued role as a forefront facility for high-angular-resolution astronomy in the era of the ELT.

Figure~\ref{fig:science} provides an overview of the spatial scales accessible with milliarcsecond angular resolution. Below, we present a non-exhaustive list of representative scientific cases.

\begin{itemize}
  \setlength\itemsep{0.2em}
  \setlength\parskip{0pt}
  \setlength\parsep{0pt}
\item \textbf{Near-Earth Asteroids (NEAs):}
Milliarcsecond-resolution imaging enables direct reconstruction of asteroid shapes, surface structures, and binarity. Such images constrain internal structure, rotational state, and surface heterogeneity, which are critical for understanding asteroid evolution and impact-risk mitigation \cite{2021A&A...654A..56V}.

\item \textbf{Trans-Neptunian Objects (TNOs):}
Direct imaging resolves the largest TNOs and binary systems, allowing measurements of shapes, sizes, and albedo variations. These observations provide key constraints on bulk density, composition, and the collisional history of the outer Solar System \cite{2014Natur.508...72B}.

\item \textbf{Small-body activity (comets and active asteroids):}
High-resolution imaging of the inner coma and jets reveals the spatial distribution of dust and gas close to the nucleus. This directly probes outgassing mechanisms and the physical processes driving activity in small bodies.

\item \textbf{Close Exoplanets:}
Interferometric imaging spatially separates close-in exoplanets from their host stars at sub-AU scales for nearby systems. Direct images enable measurements of orbital geometry, phase-dependent brightness, and rings. High spectral resolution allows characterisation of the atmosphere, presence of clouds, and chemical composition \cite{2024A&A...687A.298N}.

\item \textbf{Exomoons:}
High-contrast imaging offers a pathway to detecting and characterising large exomoons around nearby exoplanets. Spatially resolved planet--moon systems constrain satellite formation scenarios and orbital dynamics.

\item \textbf{Young Stellar Objects:}
Imaging at milliarcsecond resolution probes the inner AU-scale regions of protoplanetary disks. Direct images of gaps, spirals, and circumplanetary material reveal planet--disk interactions and constrain planet formation mechanisms and timescales \cite{2018ExA....46..517M}. At high spectral resolution, spatially resolved spectroscopy across the accretion columns, the streamers and the innermost disk directly traces mass accretion onto the star and its disk.

\item \textbf{Protoplanets and Circumplanetary disks:}
Direct imaging of circumplanetary disks provides unique constraints on mass accretion and satellite formation around young giant planets. These observations will provide direct clues on planet formation \cite{2021AJ....161..148W}.

\item \textbf{Evolved Stars:}
Interferometric imaging resolves stellar photospheres, convection cells, and dust-formation regions in the immediate circumstellar environment. These images directly constrain mass-loss processes and the shaping of planetary nebulae \cite{2021Natur.594..365M}.

\item \textbf{Stellar surface imaging:}
Direct imaging of stellar surfaces reveals spots, plages, and large-scale magnetic structures. Time-resolved images constrain stellar dynamos and magnetic activity cycles beyond the Solar analog. With high spectral resolution, resolved surface spectroscopy further disentangles rotation and pulsation, including differential rotation and non-radial pulsation modes.

\item \textbf{Stellar multiplicity and hierarchical systems:}
Milliarcsecond imaging resolves close multiple systems across the full stellar mass range. These observations constrain formation pathways, orbital evolution, and dynamical interactions in young and evolved systems \cite{2012Sci...337..444S}.

\item \textbf{Massive Stars:}
High-angular-resolution imaging resolves stellar surfaces, winds, and close environments of massive stars. Direct images reveal rotational distortion, wind clumping, and interacting binaries, providing critical input for models of massive-star evolution. At high spectral resolution, spatially resolved spectroscopy further maps the velocity field across the stellar surface and its wind, directly constraining wind kinematics and mass loss.

\item \textbf{Stellar-mass black holes:}
High-angular-resolution interferometric imaging of gravitational microlensing events enables a direct constraint on the mass distribution of stellar-mass black holes. That would provide key constraints on black hole formation channels \cite{2025ApJS..280...49M}.

\item \textbf{Galactic Centre:}
Interferometric imaging resolves the immediate environment of the Galactic Centre black hole. Time-resolved images of stars orbiting close to the black hole horizon allows tests of the general relativity and constrains the properties of Sgr~A$^\ast$.

\item \textbf{Resolved stellar populations in nearby galaxies:}
Milliarcsecond imaging separates individual stars in dense regions of nearby galaxies. This enables studies of star-formation histories, stellar evolution, and metallicity gradients. Combined with high spectral resolution, this spatial separation disentangles individual stellar spectra and provides radial velocities in addition to astrometry, within crowded fields otherwise inaccessible to single-dish spectroscopy.

\item \textbf{Supermassive black holes and binary SMBHs:}
High-resolution imaging enables the spatial separation of dual and binary SMBHs in nearby galaxies. Direct measurements of separations and orientations constrain black hole merger scenarios and gravitational-wave progenitors.

\item \textbf{Active Galactic Nuclei (AGNs):}
Interferometric imaging resolves the central parsec of AGNs, separating the dusty torus, broad-line region, and jet base. These images provide direct tests of AGN unification models and accretion--feedback coupling \cite{2020A&A...634A...1G}.

\item \textbf{Time-domain and transient phenomena:}
Time-resolved imaging of transients such as tidal disruption events, supernovae, and variable AGN structures probes the dynamical evolution of compact astrophysical systems.

\item \textbf{Measurement of the Hubble constant ($H_0$):}
High-angular-resolution imaging of geometric distance indicators, such as Cepheids, eclipsing binaries, and maser disks, enables precise and model-light distance measurements. In combination with reverberation mapping of active galactic nuclei, interferometric imaging provides direct constraints on the size and geometry of the broad-line region, anchoring luminosity distances at $z\approx 1$ and enabling an independent determination of the Hubble constant with reduced systematic uncertainties \cite{2022JHEAp..34...49A}.
\end{itemize}

\section{PROPOSED FACILITY: CONFIGURATION AND IMAGING CAPABILITY}
\label{sec:facility}

\begin{figure}
   \begin{center}
      \includegraphics[width=0.7\linewidth]{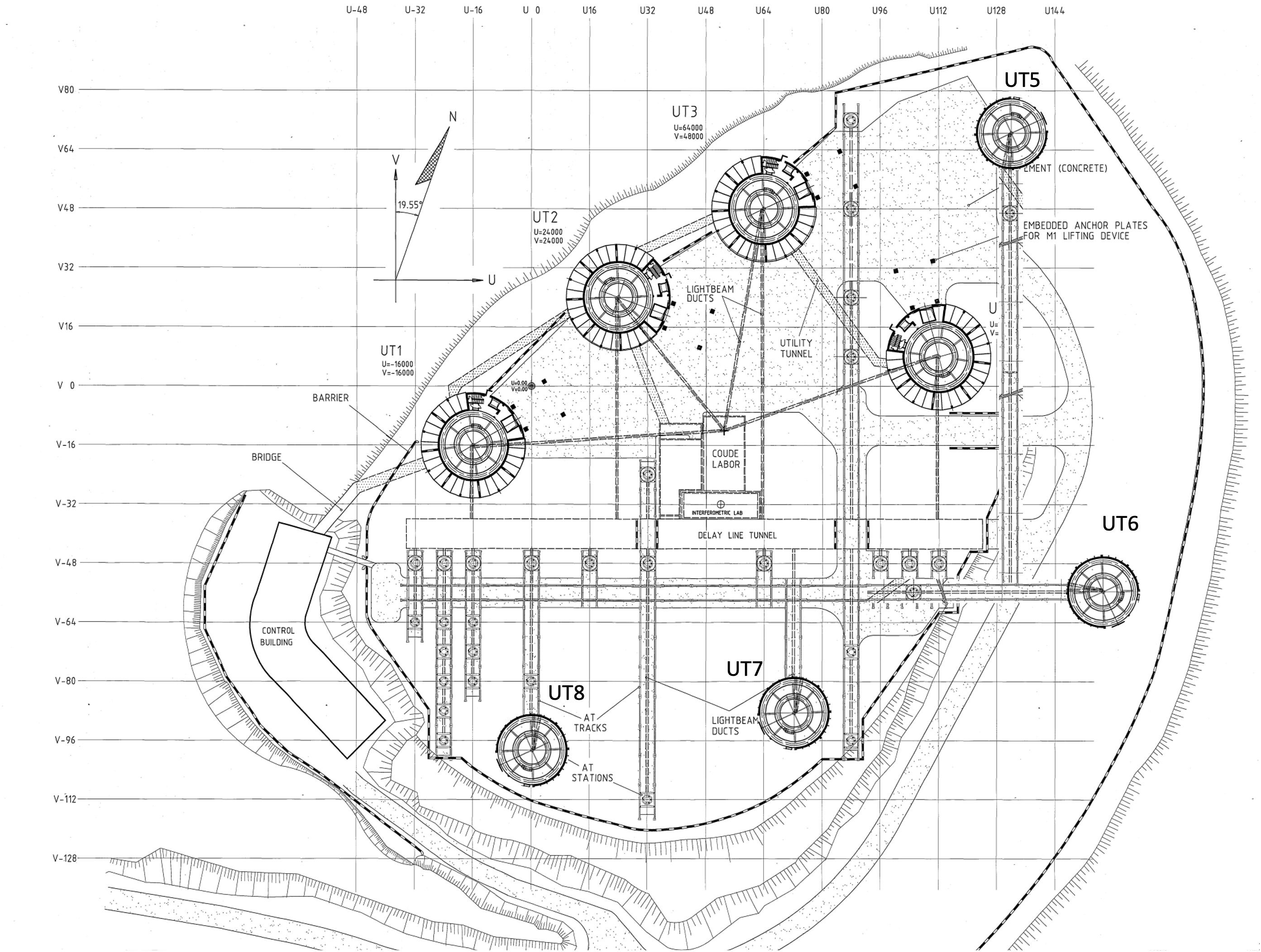}
\caption{Conceptual layout of the Paranal Observatory illustrating the proposed addition of four new 8\,m Unit Telescopes (UT5--UT8). When combined with the existing UTs, the extended array increases the number of interferometric baselines from 6 to 28, including a longest baseline of $\sim$220\,m (D2 to VST) oriented toward the north-west. The VST platform and the D2 and I1 AT stations are repurposed to accommodate the new UTs. The proposed locations are compatible with the existing delay-line tunnels, coud\'e laboratory, and observatory infrastructure.}      \label{fig:VLTI2}
   \end{center}
\end{figure}

\subsection{Siting configuration}
\label{sec:facility-siting}

With the proposed upgrade (Fig.~\ref{fig:VLTI2}), the VLTI would achieve a dense and homogeneous \textit{uv} coverage, providing baselines of up to 220\,m. This configuration would deliver angular resolutions of order 1\,mas over a field of view of approximately 50\,mas. Such enhanced \textit{uv} coverage would enable true snapshot imaging, opening the door to time-resolved studies of transient and dynamic phenomena, from near Earth asteroids to stars orbiting supermassive black holes.

We propose a specific siting configuration for the four new telescopes (Fig.~\ref{fig:VLTI2}). Two telescopes are placed on the existing D2 and I1 Auxiliary Telescope (AT) stations, which are already connected to the VLTI delay lines and therefore require no further upgrade to the observatory infrastructure. A third telescope occupies the platform currently used by the VLT Survey Telescope (VST), requiring some additional work to connect it to the UT4 light ducts. The fourth telescope is placed approximately 40\,m beyond the existing L0--M0 stations, in the direct prolongation of the existing delay lines, allowing it to be connected via a straightforward extension of the delay-line tunnel rather than the construction of a new one. This configuration keeps the site works required for three of the four telescopes minimal, and is the one adopted for the cost estimate of Sec.~\ref{sec:cost}.

\subsection{Simulated imaging capability: current 4-UT VLTI versus the proposed 8T array}
\label{sec:facility-imaging}

\begin{figure}[htp]
\begin{center}
\includegraphics[width=\linewidth]{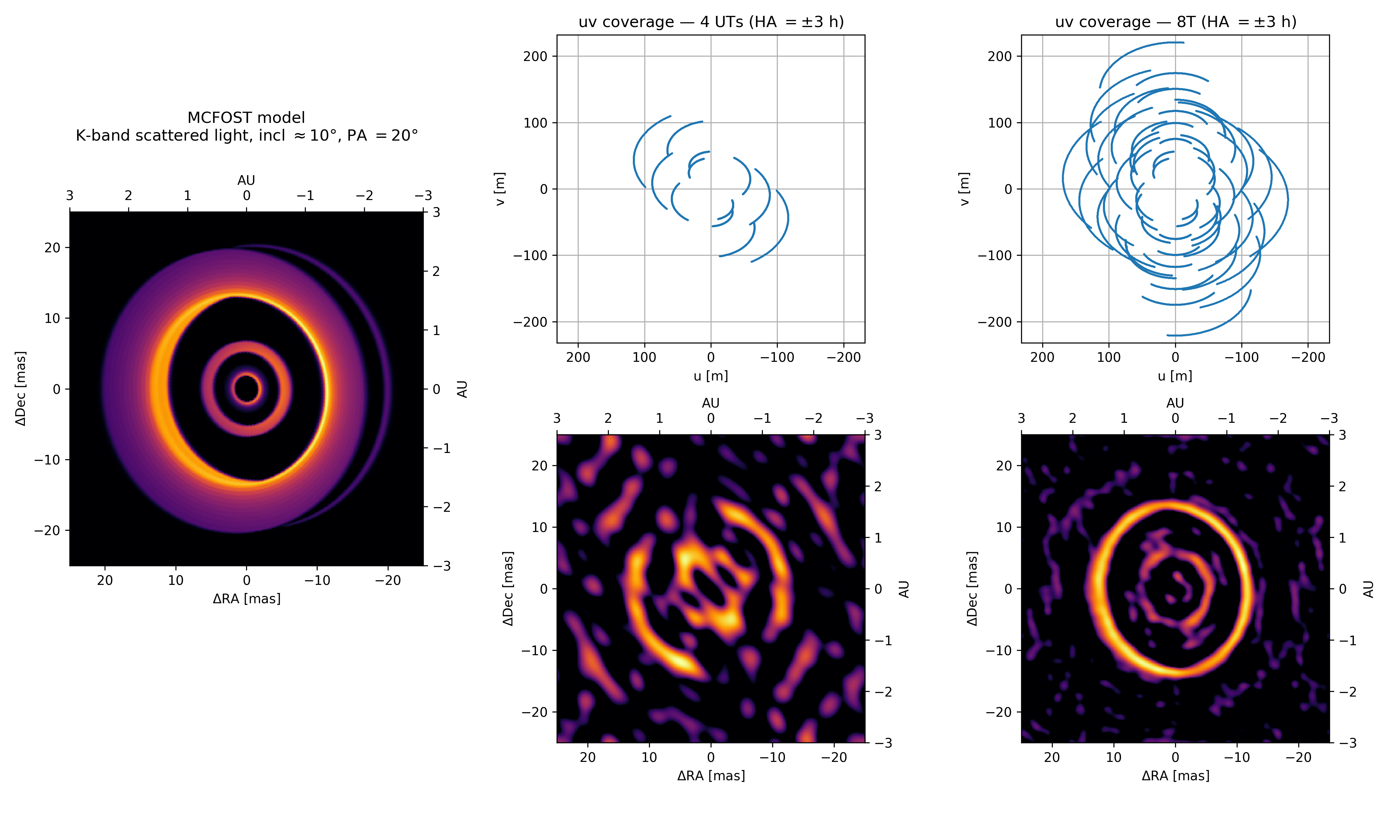}
\caption{Simulated imaging of a young stellar object (YSO) protoplanetary disk (declination $-42^\circ$, $\mathrm{HA}=\pm3$\,h). \textit{Left:} MCFOST $K$-band scattered-light model \cite{2006A&A...459..797P,2009A&A...498..967P} of a Herbig-type star (1.6\,$M_\odot$, 7200\,K) and its protoplanetary disk, with three rings at $\approx$0.3--3\,au, seen near face-on (inclination $\approx10^\circ$, position angle $20^\circ$) at $d=120$\,pc; axes in mas and au. \textit{Middle/right columns:} \textit{uv} coverage (top) and regularized reconstruction (bottom, smoothness $+$ sparsity $+$ positivity) for the current 4 UTs (6 baselines, $\approx3.3$\,mas resolution) and a hypothetical 8-telescope array (28 baselines, $\approx2.0$\,mas resolution).}
\label{fig:uvcoverage}
\end{center}
\end{figure}

To illustrate the imaging capability enabled by this \textit{uv} coverage, we simulate the reconstruction of a young stellar object (YSO) protoplanetary disk with both the current four-telescope VLTI (UT1--UT4) and the proposed eight-telescope (8T) ngVLTI array (Fig.~\ref{fig:uvcoverage}). The input model is an MCFOST radiative-transfer simulation \cite{2006A&A...459..797P,2009A&A...498..967P} of a Herbig-type pre-main-sequence star ($M=1.6\,M_\odot$, $T_{\rm eff}=7200$\,K, $R=1.5\,R_\odot$) surrounded by a disk with three bright, narrow rings at $\approx$0.3--3\,au -- a morphology reminiscent of HL~Tau, but on a much smaller physical scale -- seen nearly face-on (inclination $\approx10^\circ$, position angle $20^\circ$ east of north) and imaged in $K$-band (2.1\,\textmu m) scattered light, at an assumed distance of 120\,pc.

We simulate an observation at declination $-42^\circ$, over $\pm3$\,h of hour angle around meridian transit, sampling the spatial frequencies of the 6 baselines formed by the current array and the 28 baselines of the 8T array (each together with their Hermitian conjugates) -- corresponding to angular resolutions of $\approx3.3$\,mas and $\approx2.0$\,mas, respectively, the latter somewhat coarser than the best-case $\sim$1\,mas quoted in Sec.~\ref{sec:facility-siting} since this particular declination and hour-angle range does not access the full 200\,m baselines at all position angles. The top row of Fig.~\ref{fig:uvcoverage} shows the resulting \textit{uv}-plane coverage swept by each array as the Earth rotates -- each track is the elliptical arc traced by one baseline together with its Hermitian conjugate -- and the bottom row the corresponding image reconstructions. Direct inversion of the sparsely sampled visibilities gives a dirty image, dominated by the side lobes of the dirty beam and not directly interpretable. The regularized image reconstruction solves a single optimization problem:
\begin{equation}
\hat{x} = \operatorname*{arg\,min}_{x\,\geq\,0}
    \; \underbrace{\left\| \mathcal{S}\mathcal{F}x - V_{\rm obs} \right\|_2^2}_{\text{data fidelity}}
    \; + \; \underbrace{\lambda\,\|\nabla x\|_2^2}_{\text{smoothness}}
    \; + \; \underbrace{\mu\,\|x\|_1}_{\text{sparsity}} \, ,
\label{eq:imaging}
\end{equation}
where the three terms are, respectively, data fidelity, smoothness, and
sparsity. The symbols are defined as follows:
\begin{itemize}
  \item $x$ --- the reconstructed sky-brightness image, the unknown, with one
        non-negative value per pixel; $\hat{x}$ is the minimizer.
  \item $x \geq 0$ --- the positivity constraint: brightness cannot be
        negative.
  \item $\mathcal{F}$ --- the two-dimensional Fourier transform, mapping the
        image to the visibility (\textit{uv}) domain.
  \item $\mathcal{S}$ --- the \textit{uv}-sampling operator, a binary mask that
        retains only the spatial frequencies the array measures, together with
        the central zero-spacing point.
  \item $V_{\rm obs}$ --- the observed, sparsely sampled visibilities.
  \item $\nabla x$ --- the spatial gradient of the image; $\|\nabla x\|_2^2$
        is a Tikhonov smoothness penalty that suppresses pixel-to-pixel noise.
  \item $\lambda$ --- the weight on smoothness; larger values yield a smoother
        image.
  \item $\|x\|_1 = \sum_i |x_i|$ --- the $\ell_1$ sparsity penalty, which
        drives empty regions of the image to exactly zero.
  \item $\mu$ --- the weight on sparsity; larger values yield a darker,
        sparser background.
\end{itemize}
The positivity constraint and a fixed total-flux (zero-spacing) value are
enforced alongside Eq.~\eqref{eq:imaging}, and the problem is solved with an
accelerated, FISTA-like proximal-gradient scheme.

The current array provides only 6 independent baselines, leaving large gaps in both azimuthal and radial coverage; even after regularization, its reconstruction is dominated by dirty-beam side lobes and does not recover the ring structure. Adding the four ngVLTI telescopes at the sites identified in Sec.~\ref{sec:facility-siting} raises this to 28 baselines, and the resulting coverage is markedly denser and more azimuthally complete, filling in precisely the gaps left by the current array. The 8T reconstruction recovers all three rings with fidelity close to the input model, directly from the visibility data and without prior parametric assumptions on the disk morphology -- illustrating, in a single worked example, the imaging capability that motivates this upgrade (Sec.~\ref{sec:science}). The comparison shown here is illustrative rather than exhaustive: it fixes a single declination and hour-angle range and does not yet extend to the multi-wavelength case needed to assess the fidelity of full spectral-imaging cubes, which remains an item of ongoing work (Sec.~\ref{sec:conclusion}). Appendix~\ref{sec:appendix-uvplane} shows the underlying \textit{uv}-plane coverage of the two arrays on its own.

It should be emphasised that this reconstruction is performed at a single wavelength ($K$-band, 2.1\,\textmu m). Because spectrally dispersed instruments sample many independent wavelength channels simultaneously, the same procedure can be repeated channel by channel to build a spectral-imaging cube, in which each spatial pixel carries a full spectrum. Delivering such cubes would be a qualitatively new capability for the VLTI, bringing its imaging power close to that of ALMA and enabling the joint mapping of morphology and kinematics -- for instance, resolving the velocity field of a rotating disk or the differential structure of an emission line across a resolved source.

\section{UNIT TELESCOPE OPTICAL DESIGN}
\label{sec:optics}

\begin{SCfigure}
  \centering
  \includegraphics[width=0.45\textwidth]{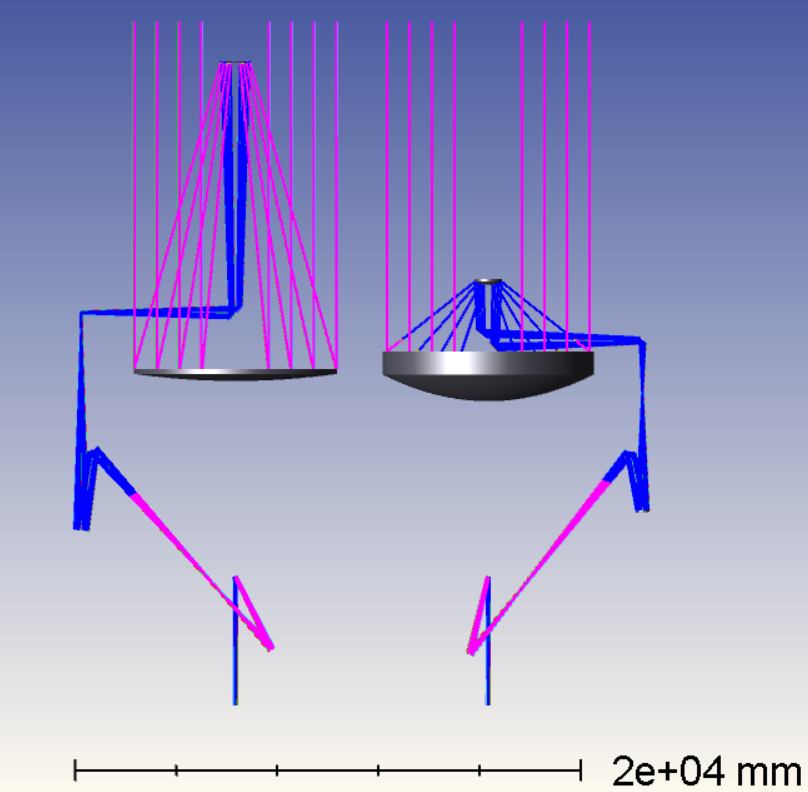}
\caption{Comparison between the current optical layout of an 8\,m Unit Telescope (UT) at Paranal (left) and the proposed compact UT concept with faster optics (right). The design preserves the same coud\'e interface as the current UT for Ground Layer Adaptive Optics (GLAO) operation, with an effective focal ratio of $f/50.28$ at the coud\'e focus, a 30\,arcsec field-of-view radius (1\,arcmin diameter), and an exit pupil located on the M8 deformable mirror (DM) at the same distance from the focal plane as in the current design (5540.2\,mm). The optical layout uses purely conic mirrors, with a segmented primary mirror composed of 60 segments of 1.2\,m each, supported on an ELT-heritage segment-support structure (see Sec.~\ref{sec:segmentation} and Fig.~\ref{fig:fea}). Using segments smaller than the ELT's 1.45\,m units allows a better fill factor of the 8\,m aperture. The M1 vertex is located at the same altitude as for the existing UTs on the Paranal platform, resulting in a coud\'e focus located 13\,044\,mm below the M1 vertex.}
\label{fig:UTdesign}
\end{SCfigure}

Delivering the collecting area, cost, and compactness required for the proposed upgrade calls for a telescope concept distinct from the existing $f/1.8$ UTs. The baseline concept is a fast ($f<1$), segmented, parabolic primary mirror (M1) that propagates the beam to a subterranean coud\'e-like focus, from which it can be relayed to an interferometric delay line and ultimately to the VLTI laboratory, in the same way as the existing UTs. The primary mirror itself is very fast, with a focal ratio of $f/0.64$; the beam is then relayed to an intermediate telescope focus at $f/10$ and finally to the coud\'e focus at $f/50.28$, with the exit pupil imaged onto the M8 deformable mirror at 5540.2\,mm from the focus, i.e.\ the same optical interface as adopted for GRAVITY+. Figure~\ref{fig:UTdesign} compares the resulting optical layout with that of the current UTs.

\begin{figure}[htp]
  \centering
  \begin{subfigure}[t]{0.48\textwidth}
    \centering
    \includegraphics[width=\textwidth]{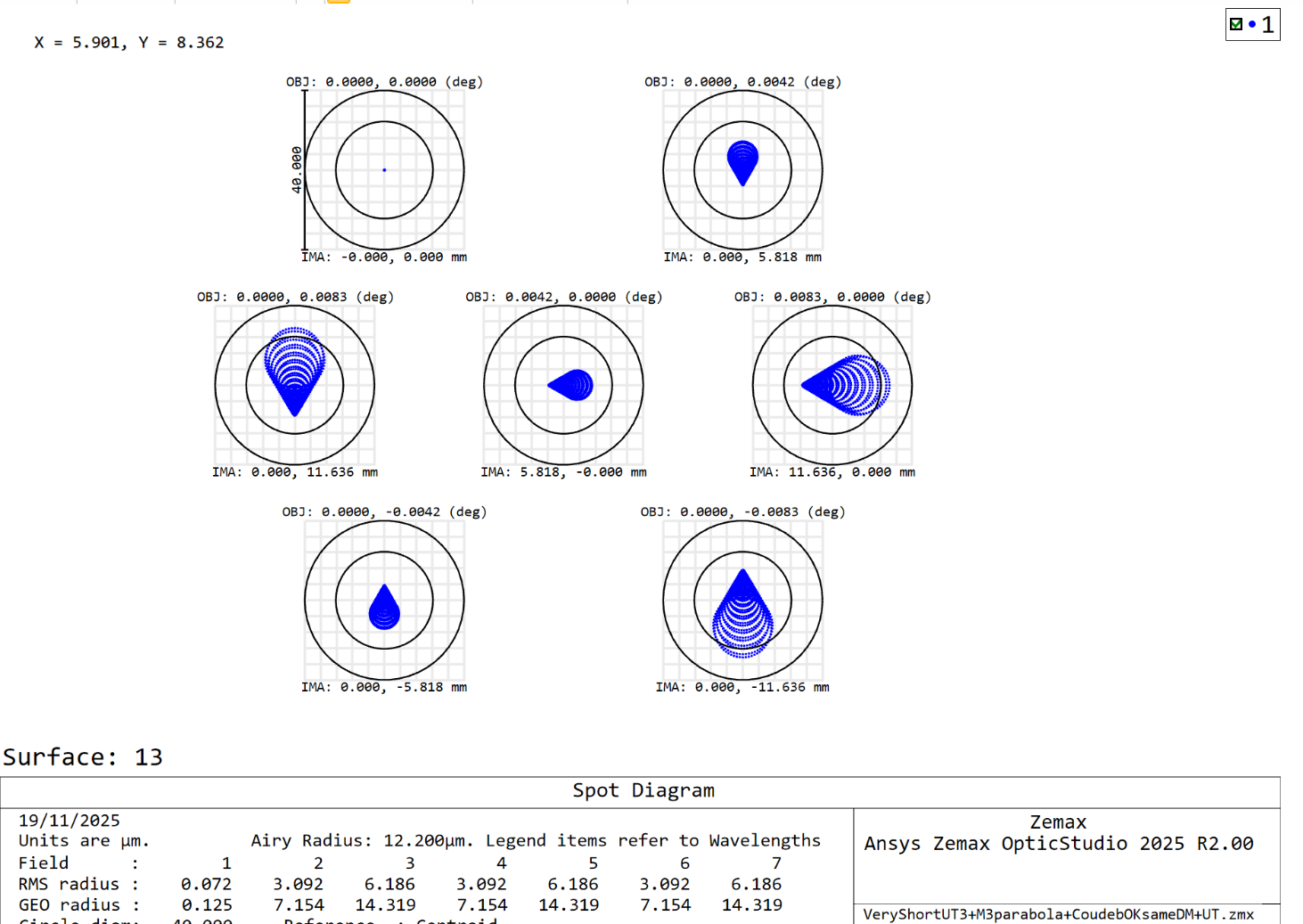}
    \caption{Telescope focus (after M4).}
  \end{subfigure}
  \hfill
  \begin{subfigure}[t]{0.48\textwidth}
    \centering
    \includegraphics[width=\textwidth]{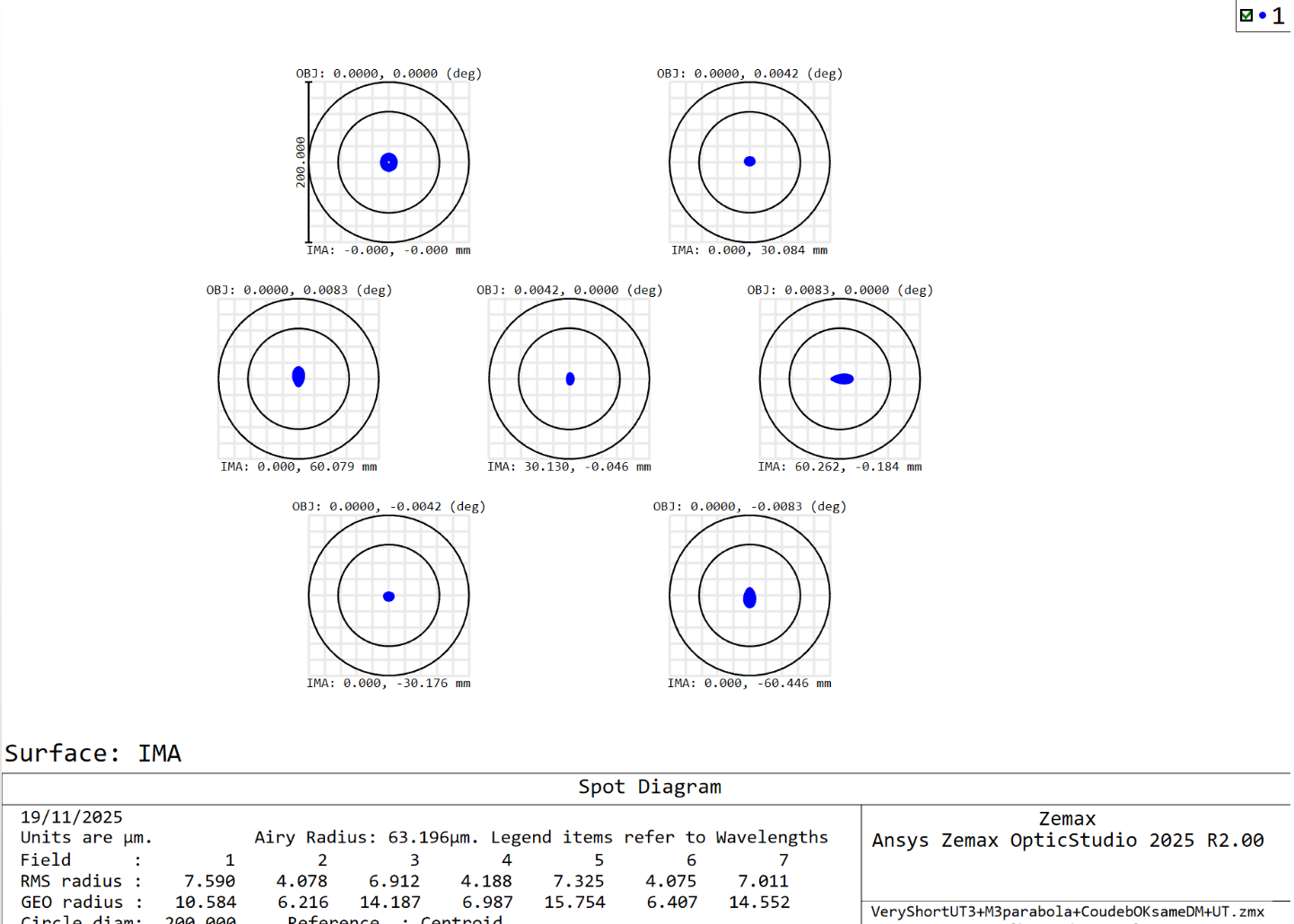}
    \caption{Coud\'e focus.}
  \end{subfigure}
  \caption{Spot diagrams across the 1\,arcmin field of view at (a) the telescope focus, immediately after M4, and (b) the final coud\'e focus, after the $f/50.28$ relay. In both cases the geometric RMS spot size remains smaller than the diffraction-limited Airy radius (12.2\,\textmu m and 63.2\,\textmu m, respectively) for all sampled field points, indicating diffraction-limited performance across the field. Ray-tracing performed with Zemax OpticStudio 2025 R2.}
  \label{fig:spots}
\end{figure}

The segmented M1 is a pure parabola; the linear field coma that this introduces off-axis is compensated further down the coud\'e relay train, which uses the same optical interface as the GRAVITY+ Adaptive Optics (GPAO) system already deployed on the existing UTs. Optical design and ray-tracing (Zemax OpticStudio) confirm that this correction is effective across the full 1\,arcmin field of view. Figure~\ref{fig:spots} shows the resulting spot diagrams at the telescope focus (immediately after the fourth fold mirror, M4) and at the final coud\'e focus, together with the diffraction-limited Airy radius at each surface. At the telescope focus, the RMS spot radius ranges from 0.07\,\textmu m on-axis to 6.2\,\textmu m at the edge of the field, well within the 12.2\,\textmu m Airy radius. At the coud\'e focus -- after the much longer effective focal length imposed by the $f/50.28$ relay -- the RMS spot radius ranges from 4.1 to 7.6\,\textmu m, again comfortably inside the 63.2\,\textmu m Airy radius. The design is therefore diffraction-limited over the full field at both surfaces. In addition, the reimaged pupil on the deformable mirror (M8) closely matches the mechanical aperture (a maximum ray radius of $\sim$50.1\,mm against a mirror radius of 55\,mm) and remains stable across all seven field points sampled, indicating negligible pupil wander with field angle.

\section{PRIMARY MIRROR SEGMENTATION: DESIGN CHOICES AND TRADE-OFFS}
\label{sec:segmentation}

The choice of an ELT-heritage segmented primary (Sec.~\ref{sec:optics}) involves a number of specific mechanical and optical trade-offs, which we summarise here.

\subsection{Segmentation pattern}

\begin{figure}[htp]
  \centering
  \begin{subfigure}[t]{0.48\textwidth}
    \centering
    \includegraphics[width=\textwidth]{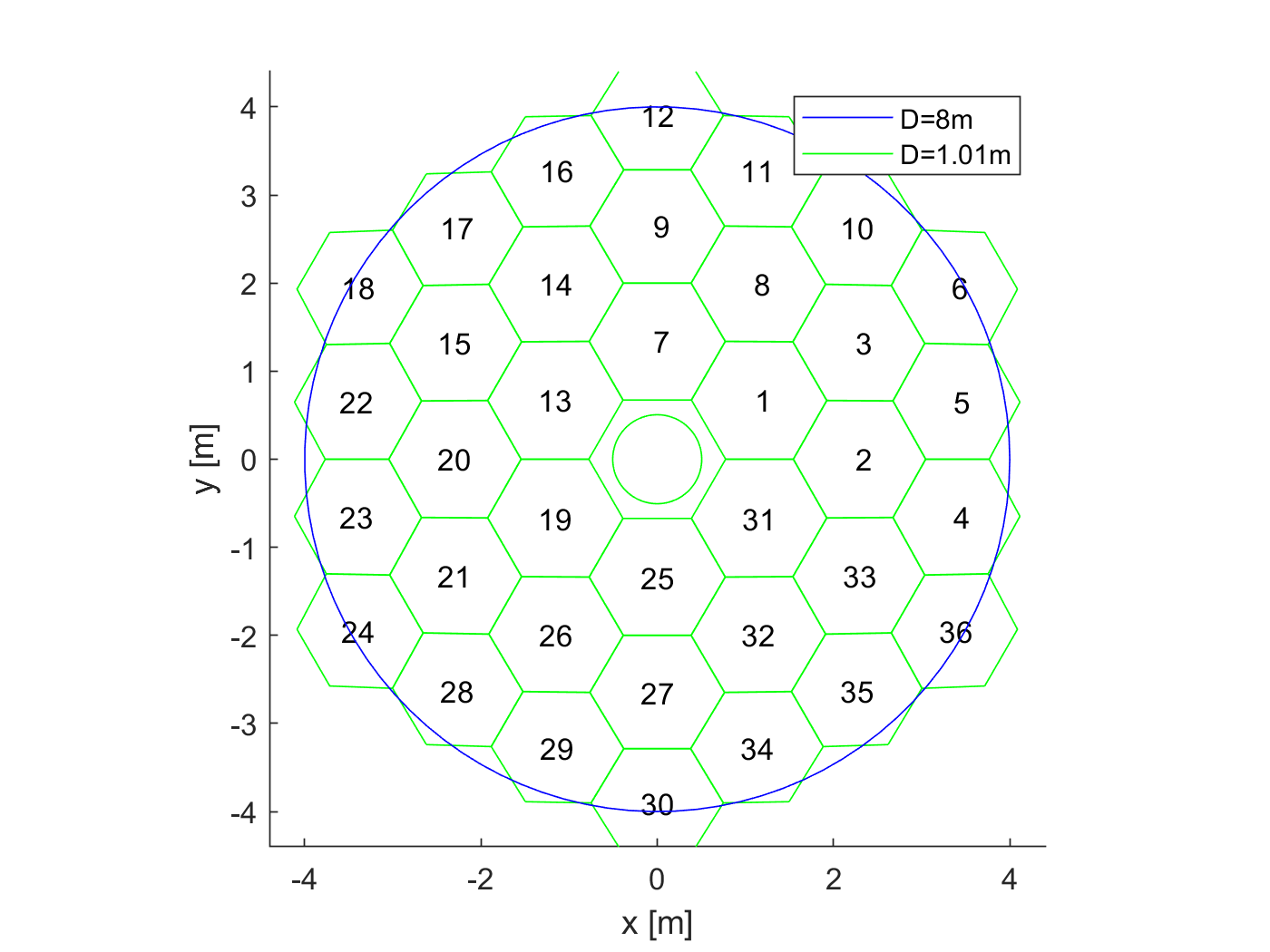}
    \caption{3 rings, 36 segments.}
  \end{subfigure}
  \hfill
  \begin{subfigure}[t]{0.48\textwidth}
    \centering
    \includegraphics[width=\textwidth]{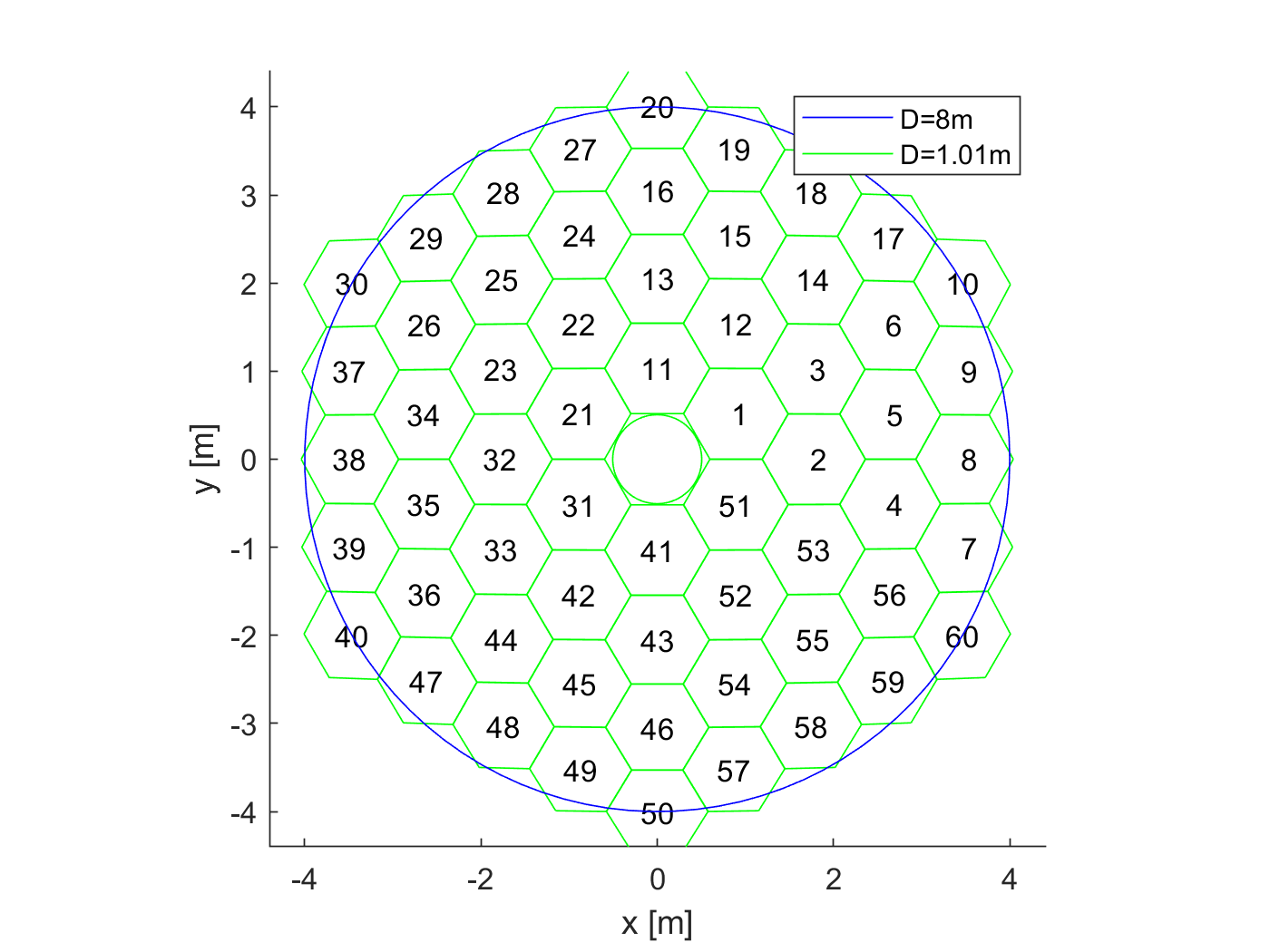}
    \caption{4 rings, 60 segments (adopted).}
  \end{subfigure}
  \caption{Candidate hexagonal segmentation patterns for the 8\,m primary mirror: (a) 3 rings of 36 segments, 1.56\,m each; (b) 4 rings of 60 segments, 1.2\,m each. The 60-segment pattern was adopted for its superior gravity print-through performance (Table~\ref{tab:segtradeoff}).}
  \label{fig:segpattern}
\end{figure}

Two candidate tilings of the 8\,m aperture were considered: a 3-ring, 36-segment pattern (1.56\,m segments), and a 4-ring, 60-segment pattern (1.2\,m segments), both built from close-packed hexagonal segments around a central obscuration (Fig.~\ref{fig:segpattern}). Hexagonal segments are strongly preferred over alternative tilings (e.g., concentric annular rings of curved trapezoidal segments) for well-established reasons: they make the most efficient use of material cut from round glass boules, present less severe corners that are easier to polish, and, owing to their six-fold symmetry, are more straightforward to support against gravity and attach to the standard three-actuator-per-segment positioning scheme (J. Nelson, ``Segmented Mirror Telescopes,'' in \textit{Optics in Astrophysics}, pp.~61--72, 2006). Because a hexagonal tiling is defined on a flat plane, projecting it onto the mirror's curved, near-parabolic surface makes each segment subtly irregular: for the 60-segment pattern, the corner-to-corner width of individual segments varies by up to $\sim$4\% (from 1.15 to 1.20\,m) depending on ring number, a well-known consequence of tiling a curved surface that must be accounted for in the segment procurement plan, since no two segments are then perfectly identical. A more exotic alternative, based on a Goldberg-polyhedron-like decomposition of the best-fitting sphere using a small number of pentagonal elements to reduce the number of unique segment shapes, was also considered, but was not pursued pending further study of its impact on the diffraction pattern and on segment-to-segment regularity.

\subsection{Segment size, thickness, and gravity print-through}
\label{sec:seg-printthrough}

The 60-segment pattern was adopted over the 36-segment alternative primarily because smaller segments are markedly stiffer against gravity print-through -- the residual, intrinsic figure error of an individual segment under its own weight, which, unlike the system-level flexure discussed in Sec.~\ref{sec:flexure}, cannot be corrected by rigid-body realignment and must instead be controlled by the segment support design itself. Following the classical plate-deflection formula of Nelson (Keck Observatory technical report, 1982) for a mirror segment on discrete point supports,
\begin{equation}
\delta_{\rm rms} = k\,\frac{q\,a^4}{D}\,, \qquad D = \frac{E h^3}{12(1-\nu^2)}\,,
\label{eq:printthrough}
\end{equation}
where $a$ is the segment radius, $h$ its thickness, $q$ the gravitational load per unit area, $E$ and $\nu$ the material's elastic modulus and Poisson ratio (for Zerodur, $E=90$\,GPa, $\nu=0.24$, $\rho=2540\,{\rm kg/m^3}$), and $k$ a dimensionless factor set by the support-point topology, the RMS print-through scales steeply with segment size ($a^4$) and thickness ($h^{-3}$). Table~\ref{tab:segtradeoff} illustrates this trade-off for both segmentation patterns and a range of segment thicknesses, evaluated for an 18-point and a 27-point whiffletree support. Reducing the segment size from 1.56 to 1.2\,m alone should reduce the print-through by a factor of $(1.2/1.56)^4\approx0.35$; the tabulated values (e.g., 8.0 vs.\ 22.9\,nm at $h=50$\,mm) closely match this scaling.

\begin{table}[ht]
\caption{Trade-off between segmentation pattern, segment thickness $h$, gravity print-through $\delta_{\rm rms}$ (18- and 27-point whiffletree support), and mass. $m_{\rm seg}$ is the total mass (Zerodur and support) of a single segment; $M_{\rm M1}$ is the total primary-mirror mass. The adopted baseline is shown in bold.}
\label{tab:segtradeoff}
\begin{center}
\begin{tabular}{|c|c|c|c|c|c|c|}
\hline
\rule[-1ex]{0pt}{3.5ex} $n_{\rm seg}$ & $d$ [mm] & $h$ [mm] & $\delta_{18}$ [nm] & $\delta_{27}$ [nm] & $m_{\rm seg}$ [kg] & $M_{\rm M1}$ [kg] \\
\hline
\rule[-1ex]{0pt}{3.5ex} 36 & 1560 & 50 & 22.9 & 10.1 & 389 & 14\,007 \\
\hline
\rule[-1ex]{0pt}{3.5ex} 36 & 1560 & 40 & 35.8 & 15.8 & 339 & 12\,193 \\
\hline
\rule[-1ex]{0pt}{3.5ex} 60 & 1200 & 50 & 8.0 & 3.6 & 270 & 16\,182 \\
\hline
\rule[-1ex]{0pt}{3.5ex} 60 & 1200 & 40 & 12.5 & 5.5 & 240 & 14\,392 \\
\hline
\rule[-1ex]{0pt}{3.5ex} \textbf{60} & \textbf{1200} & \textbf{35} & \textbf{16.4} & \textbf{7.2} & \textbf{225} & \textbf{13\,498} \\
\hline
\end{tabular}
\end{center}
\end{table}

The adopted baseline -- 60 segments of 1.2\,m and 35\,mm thickness -- achieves an RMS print-through of 16.4\,nm (18-point support) or 7.2\,nm (27-point support), essentially matching the print-through performance of the ELT's own 1.45\,m, 50\,mm segments (17.2\,nm and 7.6\,nm, respectively, using the same formalism) despite using 30\% less glass thickness: the smaller segment size compensates for the reduced stiffness of a thinner mirror. The choice between 18- and 27-point support directly trades mechanical complexity (and hence cost, Sec.~\ref{sec:cost}) against optical quality: a dedicated point-support trade-off study for ELT-class segments found that increasing the support-point count from 18 to 27 typically improves the RMS surface error by a factor of $\sim$2, but that not every high-point-count geometry is mechanically feasible -- e.g., two of three corner-centred 27-point whiffletree configurations studied were found to suffer from collinear-point singularities or actuator-tripod interference (CESA/MEDIA Consultores de Ingenier\'ia, ELT M1 segment-support trade-off study, Ref.~E-CES-TRE-189-0006). We adopt the 18-point, side-centred configuration as our baseline, noting that upgrading to 27 points remains an available performance margin should it be required.

\subsection{Structural feasibility}
\label{sec:seg-structural}

\begin{figure}[htp]
  \centering
  \includegraphics[width=0.75\textwidth]{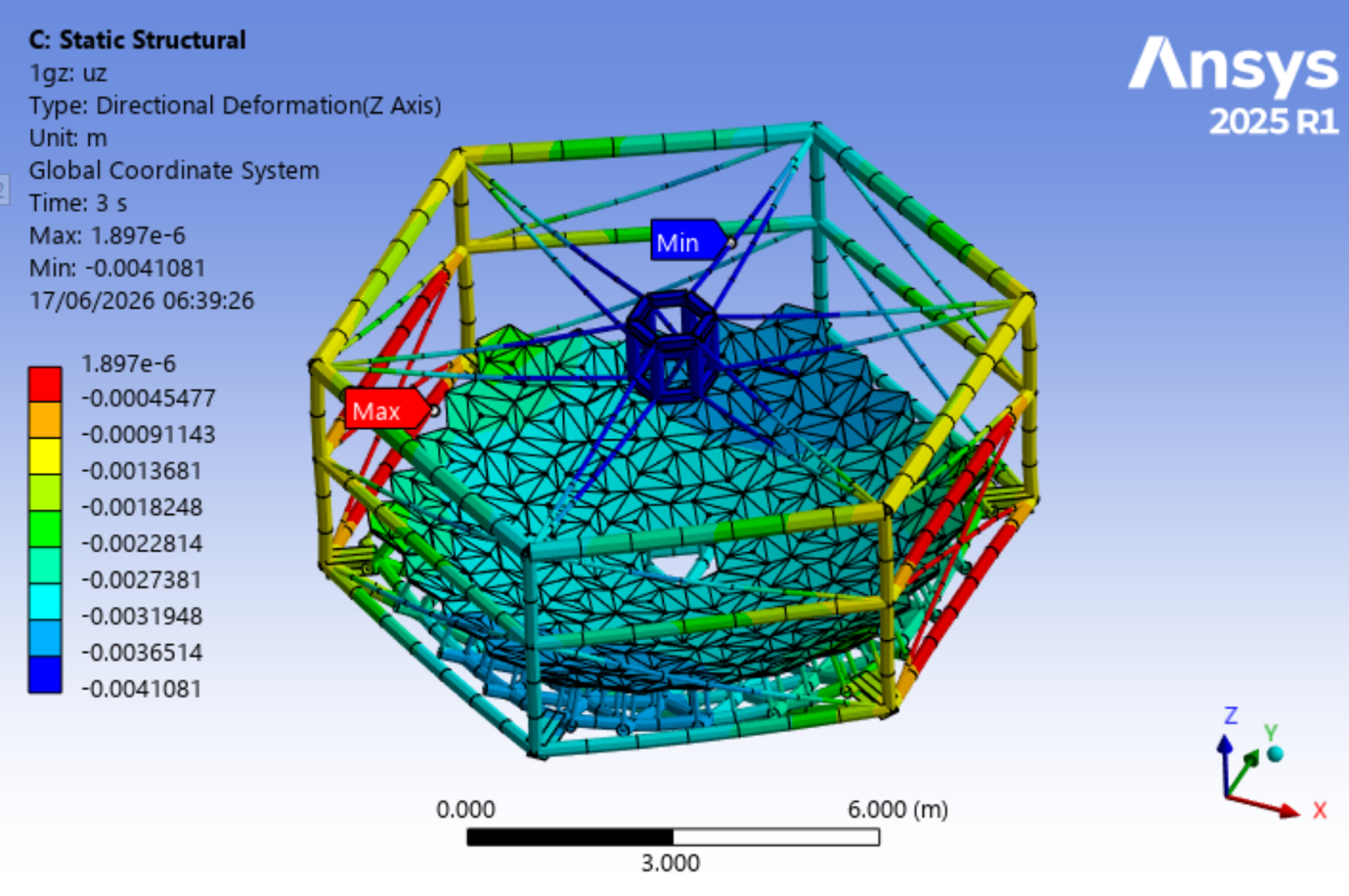}
  \caption{Finite-element (ANSYS Static Structural) analysis of the primary-mirror segment-support truss under self-weight (gravity load along the telescope $z$-axis). The colour scale shows the directional deformation along $z$, ranging from negligible ($\sim$2\,\textmu m) near the central hub to a maximum of $-4.1$\,mm at the outer edge of the structure, consistent with a stiff, ELT-heritage space-frame design at the 8\,m scale.}
  \label{fig:fea}
\end{figure}

A finite-element model of the complete 60-segment primary, including its ELT-heritage support cell (Fig.~\ref{fig:fea}), gives a total altitude moving mass of 37.1\,tons, adopted directly as the altitude-structure allocation used in the cost estimate of Sec.~\ref{sec:cost}. Under self-weight, the model shows deflections of at most a few millimetres at the extremities of the support truss, supporting the assumption that an ELT-derived design is structurally adequate at the 8\,m scale.

\subsection{Off-axis segment fabrication}

Because the primary is a fast, parabolic surface (Sec.~\ref{sec:optics}), the polishing figure required for each segment departs from a simple sphere by an amount that grows with distance from the optical axis; outer-ring segments therefore require increasingly custom, non-replicable aspheric figures, a well-documented cost driver in segmented-mirror fabrication. A useful benchmark is the comparison between Keck (0.9\,m segments) and the CELT design study (a precursor to the Thirty Meter Telescope, 0.5\,m segments): the outermost-segment departure from best-fit sphere is $C_{20}\approx-100\,\mu$m for Keck against only $\approx-19\,\mu$m for CELT (J. Nelson, ``Primary Mirror Segment Fabrication for CELT,'' 2000), a direct consequence of CELT's smaller segments. The same qualitative trend was confirmed for our own 60-segment array using the same formalism: the required best-fit-sphere curvature changes systematically from the innermost to the outermost sampled segment, confirming that outer-ring segments require a progressively more aspheric, less replicable polishing figure -- an expected cost driver that favours the smaller, 1.2\,m segments adopted here over the 36-segment alternative, consistent with the segment-cost assumption already used in Table~\ref{tab:cost}.

\section{GRAVITATIONAL FLEXURE AND OPTICAL SENSITIVITY}
\label{sec:flexure}

Beyond the segment-level print-through discussed in Sec.~\ref{sec:seg-printthrough} -- which is intrinsic to each segment's own support and cannot be corrected by rigid-body realignment -- the telescope structure as a whole flexes under gravity as it tracks in elevation, changing the relative alignment of M1 and M2. Because this system-level flexure is, in principle, correctable by an actively controlled M2 hexapod, as on the existing UTs, it constitutes a different kind of budget: not a fundamental image-quality limit, but a requirement on the stroke, precision, and update rate of the active-alignment system. We quantify this requirement using a dedicated finite-element and optical-sensitivity model, distinct from the static, as-built optical performance already established in Sec.~\ref{sec:optics}.

\subsection{Finite-element model and rigid-body flexure}

The finite-element model of Sec.~\ref{sec:seg-structural} was used to compute the rigid-body decentre and tilt of M1 and M2 under three orthogonal unit-gravity load cases (gravity along each of the telescope's $x$, $y$, and $z$ axes). Because an alt-azimuth telescope's orientation relative to gravity depends only on the zenith distance angle (ZA), these three load cases can be linearly combined to give the flexure at any ZA relative to a zenith-pointing ($\mathrm{ZA}=0^\circ$) reference, at which the telescope is assumed to be calibrated. In practice the nominal elevation-dependent deflection is produced almost entirely by the $y$ and $z$ load cases; the $x$ load case contributes only marginally -- e.g.\ through a small misalignment of the azimuth axis with respect to gravity -- so that the flexure is, to good approximation, a combination of the two dominant cases.  At $\mathrm{ZA}=90^\circ$ (pointing at the horizon), the most demanding case relative to this reference, the model predicts a relative M2-to-M1 misalignment (Table~\ref{tab:wfe}) of approximately 870\,\textmu m of axial spacing (piston), 670\,\textmu m of lateral decentre, and 58\,\textmu rad of tilt, dominated by the $\sim$3.6\,m axial separation between M1 and M2 acting as a lever arm on any differential structural rotation.

\subsection{Wavefront error budget}

These rigid-body motions were propagated into wavefront error using Zernike sensitivity coefficients derived from optical ray-tracing (Sec.~\ref{sec:optics}): the response of each low-order Zernike mode to a unit decentre or tilt of M2 relative to M1. Because the flexure at any ZA is itself a linear combination of the same two orthogonal unit-gravity load cases used throughout this section, both the wavefront-error amplitude and its instantaneous rate of change at intermediate pointings follow directly from the same closed-form combination (the rate being simply its analytic derivative with respect to ZA, scaled by the assumed tracking rate). Table~\ref{tab:wfe} reports the resulting raw, uncorrected amplitude and rate of change at $\mathrm{ZA}=10^\circ$, $45^\circ$, and $90^\circ$ relative to the zenith-pointing reference, assuming a representative elevation-tracking rate of 10$^\circ$/hour (piston, tip, tilt, and trefoil terms are omitted as optically benign or negligible). In table\,\ref{tab:wfe} we observe a `typical' behaviour of the aberrations with changing elevation. Physically, this reflects the two different gravity components at play: focus and spherical aberration are driven predominantly by the gravity component along the optical ($z$) axis, which grows as $(1-\cos\mathrm{ZA})$ and therefore rise together as the telescope moves from zenith toward the horizon, whereas coma is driven by the lateral decentre between M1 and M2 induced by the gravity component perpendicular to the optical axis, which instead grows as $\sin\mathrm{ZA}$. 

\begin{table}[ht]
\caption{Raw, uncorrected system-level gravitational flexure at zenith distance angles 10$^\circ$, 45$^\circ$, and 90$^\circ$, relative to a zenith-pointing reference. The upper block gives the rigid-body displacement of M2 relative to M1 (translations in \textmu m, tilt in \textmu rad); the lower block gives the low-order Zernike wavefront error these motions induce. The Rate columns give the instantaneous rate of change for an assumed 10$^\circ$/hour elevation-tracking rate, in the unit of the corresponding amplitude per 5\,min, obtained as the finite-difference derivative of the tabulated flexure with respect to ZA. The remaining rigid-body degrees of freedom ($\Delta x\lesssim2$\,\textmu m, and rotations about the $y$ and $z$ axes $\lesssim1$\,\textmu rad) are negligible and omitted; the ZA=45$^\circ$ displacement values are interpolated between the 40$^\circ$ and 50$^\circ$ finite-element samples.}
\label{tab:wfe}
\begin{center}
\small
\begin{tabular}{|l|c|c|c|c|c|c|c|}
\hline
\rule[-1ex]{0pt}{3.5ex} \multirow{2}{*}{Quantity} & \multirow{2}{*}{Term} & \multicolumn{3}{c|}{Amplitude} & \multicolumn{3}{c|}{Rate [/5\,min]} \\
\cline{3-8}
\rule[-1ex]{0pt}{3.5ex} & & ZA=10$^\circ$ & ZA=45$^\circ$ & ZA=90$^\circ$ & ZA=10$^\circ$ & ZA=45$^\circ$ & ZA=90$^\circ$ \\
\hline
\hline
\multicolumn{8}{|l|}{\rule[-1ex]{0pt}{3.5ex}\textit{Rigid-body displacement of M2 relative to M1}} \\
\hline
\rule[-1ex]{0pt}{3.5ex} Axial spacing (piston) & $\Delta z$ [\textmu m] & 13 & 258 & 873 & 2.2 & 9.0 & 12.6 \\
\hline
\rule[-1ex]{0pt}{3.5ex} Lateral decentre & $\Delta y$ [\textmu m] & 110 & 456 & 667 & 9.1 & 6.8 & 1.3 \\
\hline
\rule[-1ex]{0pt}{3.5ex} Relative tilt & $\theta_x$ [\textmu rad] & 10 & 41 & 58 & 0.8 & 0.6 & 0.1 \\
\hline
\hline
\multicolumn{8}{|l|}{\rule[-1ex]{0pt}{3.5ex}\textit{Induced wavefront error (Zernike, [\textmu m])}} \\
\hline
\rule[-1ex]{0pt}{3.5ex} Focus & $Z_4$ & 1 & 28 & 95 & 0.1 & 1.0 & 1.4 \\
\hline
\rule[-1ex]{0pt}{3.5ex} Astigmatism & $Z_5$ & 0.0 & 0.2 & 0.3 & 0.0 & 0.0 & 0.0 \\
\hline
\rule[-1ex]{0pt}{3.5ex} Coma & $Z_8$ & 30 & 123 & 174 & 2.5 & 1.8 & 0.0 \\
\hline
\rule[-1ex]{0pt}{3.5ex} Spherical aberration & $Z_{11}$ & 1 & 15 & 51 & 0.1 & 0.5 & 0.7 \\
\hline
\end{tabular}
\end{center}
\end{table} 

As expected for a two-mirror system, the dominant terms are focus (up to 95\,\textmu m at $\mathrm{ZA}=90^\circ$, driven by the axial spacing change) and coma (up to 174\,\textmu m, driven by the relative tilt), both of which lie within the correction range of a standard five-degree-of-freedom M2 hexapod (decentre in $x,y$; tilt about $x,y$; piston along $z$) -- the same actuator architecture already used operationally on the existing UTs, following an elevation-dependent alignment look-up table. A smaller but non-negligible spherical-aberration term (up to 51\,\textmu m) arises from the same axial-spacing change and is expected to be substantially reduced by the same piston correction that restores nominal focus, since both terms share this common physical origin. Astigmatism remains comparatively small (up to 0.3\,\textmu m) across the full elevation range. These results should be read as a raw sensitivity budget -- sizing the stroke and precision required of the M2 active-alignment system -- rather than as a post-correction, delivered-image-quality budget; closing the loop with a specific hexapod control model is left for future work.

\subsection{Update-rate requirement and DM offloading}

The rate of change of the flexure, not its absolute amplitude, is what ultimately sets the required correction bandwidth. In closed-loop operation, low-order aberrations are first sensed and corrected by the deformable mirror (DM) of the adaptive optics system, located at the telescope's final exit pupil on M8 (Sec.~\ref{sec:optics}). For the GRAVITY+ Adaptive Optics (GPAO) system already baselined for the existing UTs, and adopted here as the coud\'e-train interface for the new telescopes, this is an ALPAO $43\times43$ voice-coil deformable mirror with 1432 actuators on a 2.62\,mm pitch, specified for a local (high-order) stroke of order 20\,\textmu m peak-to-valley wavefront per group of $3\times3$ actuators \cite{bourdarot2024gpao,gravityplus2026gpao}. Because this stroke budget is small compared to the hundreds of microns of focus and coma predicted in Table~\ref{tab:wfe}, the DM alone cannot absorb the full gravitational flexure over an elevation track without saturating: the low-order component of the wavefront error must instead be continuously \emph{offloaded} from the DM onto slower, larger-stroke actuators upstream -- primarily the M2 hexapod (Sec.~\ref{sec:flexure}), and, on longer timescales, the active-optics and pointing model acting through M1 and the telescope structure itself.

GPAO already implements exactly this kind of offload architecture operationally, which we use to size the requirement for our concept. Its control system includes a dedicated secondary loop that offloads the DM shape onto the M1 active-optics system; on the existing, comparatively stiff $f/1.8$ UTs, this loop has in practice been decommissioned, because the DM alone was found to have enough stroke to absorb the (much smaller) static and quasi-static aberrations of those telescopes \cite{gravityplus2026gpao}. A second secondary loop offloads the mean tip-tilt component of the DM shape onto the telescope pointing axes every 3\,s, with a servo gain of $\approx$0.5; a faster, $\sim$100\,Hz offload directly to the M2 rapid-guiding link was considered during the design phase but discarded, since it was found to inject M2 mount noise back into GPAO's own correction bandwidth \cite{gravityplus2026gpao}.

Given that our concept predicts a raw flexure one to two orders of magnitude larger than that of the existing UTs (hundreds of microns, Table~\ref{tab:wfe}, against a DM stroke of order tens of microns), the M1-offload loop that GPAO's existing UTs can safely leave decommissioned would very plausibly need to be re-enabled for the proposed telescopes -- the control infrastructure for it already exists within the GPAO design, requiring no new development, only a revised offload gain and cadence. This cadence is set by the rate of change in Table~\ref{tab:wfe}, not by the total amplitude: the rates are largest near zenith (up to 2.5\,\textmu m per 5\,min for coma) and fall toward the horizon, except for focus and spherical aberration, whose rate keep increasing to 1.4\,\textmu m per 5\,min for focus and 0.74\,\textmu m per 5\,min for spherical aberration at $\mathrm{ZA}=90^\circ$. An offload period of order tens of seconds to a minute -- comfortably faster than GPAO's own demonstrated 3\,s tip-tilt offload cadence -- would keep the DM's own contribution to the flexure well within its local stroke budget at all pointings, leaving correction of the bulk, low-frequency amplitude to the M2 hexapod and, through the reactivated M1 offload, to the telescope's active-optics system, following the same architecture already validated operationally by GPAO.

\subsection{M1 versus M2 offload: mount noise and scalloping}

The choice of \emph{where} to offload the bulk correction -- M2 or M1 -- is not neutral for an interferometer. Offloading onto the M2 hexapod is optically exact, since M2 is a single continuous mirror, but it requires physically moving M2's mount at whatever cadence the offload demands; any dynamical content in that motion couples directly into the optical path length of the beam relayed to the VLTI beam-combination laboratory, degrading fringe stability in a way that a single-dish AO system would not suffer from. This is precisely the concern that led GPAO to discard its own fast, $\sim$100\,Hz M2 offload in favour of a slower channel (Sec.~\ref{sec:flexure}). Offloading instead onto the segmented M1 -- adjusting the piston, tip, and tilt of the 60 segments to track the target low-order shape -- keeps the correction upstream of the coud\'e relay and avoids exciting the M2 mount altogether, at the cost of only ever approximating a smooth aberration with 60 flat-moving, rigid facets.

This approximation is not free: a segment's three rigid-body degrees of freedom can null the local mean value and mean slope of the target wavefront at that segment, but not its local curvature, since rigid-body motion has no second-order term to offer. The uncorrected local curvature is left behind as a residual, periodic pattern at the segment scale, commonly referred to as \emph{scalloping} -- a distinctive ``staircase'' approximation of the continuous parent aberration by the array of rigid facets. For a mirror of diameter $D$ tiled by $N$ segments of diameter $2a\ll D$, this residual scales down with segment count approximately as $(2a/D)^2\sim 1/N$: for our 60-segment array, this suggests only a few percent of the original aberration amplitude survives as scalloping once each segment has been individually aligned, though we emphasise this is an order-of-magnitude scaling law rather than an exact result, and a full evaluation for the specific gravity-flexure amplitudes of Table~\ref{tab:wfe} is left for future work. In practice, this suggests that M1 offload is best suited to removing the bulk, quasi-static part of the flexure -- for which even a percent-level scalloping residual is a large absolute improvement -- while the M2 hexapod, free of any scalloping penalty, remains the more exact (if noise-sensitive) actuator of last resort for large amplitude corrections. The two are not mutually exclusive: the same offload architecture already validated by GPAO's decommissioned M1 loop and retained M2 tip-tilt loop shows that both channels can coexist, each handling the part of the correction it is best suited for.

\section{COST AND FEASIBILITY ASSESSMENT}
\label{sec:cost}

\subsection{Scope and assumptions}
\label{sec:cost-scope}

We provide a back-of-the-envelope estimate of the cost of a single fast ($f<1$) segmented 8\,m telescope of the type described in Sec.~\ref{sec:optics}, and use it to derive an order-of-magnitude cost for the four-telescope upgrade. The estimate assumes the siting configuration described in Sec.~\ref{sec:facility} -- two telescopes on existing AT stations (D2, I1), one on the VST platform, and one in the prolongation of the L0--M0 delay-line stations -- so that no other change to the VLTI infrastructure (coud\'e laboratory, VLTI beam-combination laboratory) is required beyond the VST light-duct connection and the delay-line tunnel extension already noted. Under this assumption, the estimate below covers the telescope unit only -- primary segments, secondary, coud\'e relay optics, mount, enclosure, and local ancillary instrumentation -- and does not include the cost of the VST light-duct connection, the delay-line tunnel extension, any further VLTI-wide infrastructure or instrumentation upgrade, or project-level costs such as system engineering, integration, and operations. All figures are order-of-magnitude and expressed in 2026 Euro unless stated otherwise.

\subsection{Historical reference costs}
\label{sec:cost-history}

Historical costs for comparable facilities, drawn from the publicly available literature, provide a useful benchmark. The two Keck telescopes cost 90\,M\$ each (1991). The four UTs of the VLT cost a combined 664\,M\,DM (1998), including infrastructure, or approximately 150\,M\,DM (70\,M\texteuro) per telescope. The two Gemini telescopes cost 190\,M\$ (1999). Using an approximate inflation factor of 1.5 between 2000 and 2026. The weight of a conventional $f/1.8$ 8\,m-class telescope -- itself a good proxy for cost -- is typically 300 to 500\,tons. In general, a multi-purpose 8-10 metre class f/1.8 primary telescope, inflation adjusted should cost of order 110 to 120 M\texteuro.  

\subsection{Cost breakdown}

The proposed concept uses 60 segments of 1.2\,m on an ELT-heritage support structure (Sec.~\ref{sec:segmentation}), giving a telescope structure markedly lighter than a conventional $f/1.8$ UT. We budget a total moving mass of 150\,tons: 50\,tons for the altitude structure, and approximately three times that for the azimuth structure. The 50-ton altitude budget carries margin above the 37.1\,tons given by the bare finite-element model of Sec.~\ref{sec:seg-structural}, which covers only the M1 segments and their mechanical support; the difference allows for rotation bearings, cabling, and other ancillary hardware not represented in the model shown in Fig.~\ref{fig:fea}. This 150-ton total is achieved because the coud\'e-only optical path dispenses with a Nasmyth platform -- a comparatively heavy and bulky structure on conventional UTs -- and it sits roughly a factor of two to three below the 300--500\,ton range quoted above for existing $f/1.8$ 8\,m telescopes. This reduction in moving mass is the principal structural lever behind the cost reduction derived below. Table~\ref{tab:cost} itemises the cost of a single telescope. Each segment (glass, polishing, support, and actuation) is significantly lighter and smaller than the ELT segments but can use the edge sensors and actuators designed for the ELT. We have costed a completed segment at 300,000\,\texteuro; including a spare set, the segment procurement totals approximately 20\,M\texteuro. An adaptive secondary mirror (ASM) -- the most expensive, but most capable, option for correcting the telescope's optical aberrations -- is costed at 10\,M\texteuro. The coud\'e relay train, comprising seven flat mirrors of order 1\,m each, is costed at 5\,M\texteuro\ for the set. Only one side will be equipped with a coude train and the M3 is adjustable for pointing but not on a rotating stage. The telescope structure (mount) is costed on a per-kilogram basis of 100,000\,\texteuro\ per telescope-ton (excluding the enclosure), the estimated 150-ton structure costs 15\,M\texteuro. The enclosure and foundations, benefiting from the telescope's reduced size and the absence of instrumentation supporting structures, are capped at 10\,M\texteuro; since the design routes the beam directly to the coud\'e focus and dispenses with a Nasmyth focus, the dome need only clear the telescope tube itself, rather than the wider swept volume required by a Nasmyth instrument platform, keeping this cost contained. Finally, all ancillary equipment in the coud\'e room -- guiding, star separation, and wavefront sensing -- is capped at 10\,M\texteuro.

\begin{table}[ht]
\caption{Back-of-the-envelope cost breakdown for a single fast ($f<1$), 8\,m-class segmented Unit Telescope (2026 Euro).}
\label{tab:cost}
\begin{center}
\begin{tabular}{|l|c|}
\hline
\rule[-1ex]{0pt}{3.5ex} Item & Cost (M\texteuro) \\
\hline
\rule[-1ex]{0pt}{3.5ex} Primary mirror segments (60 $\times$ 1.2\,m, incl.\ spare set) & 20 \\
\hline
\rule[-1ex]{0pt}{3.5ex} Adaptive secondary mirror & 10 \\
\hline
\rule[-1ex]{0pt}{3.5ex} Coud\'e relay train (7 flat mirrors and supports) & 5 \\
\hline
\rule[-1ex]{0pt}{3.5ex} Telescope structure (mount) & 15 \\
\hline
\rule[-1ex]{0pt}{3.5ex} Enclosure and foundations & 10 \\
\hline
\rule[-1ex]{0pt}{3.5ex} Ancillary instrumentation (guiding, star separator, WFS) & 10 \\
\hline
\rule[-1ex]{0pt}{3.5ex} contingency  & 10 \\
\hline
\rule[-1ex]{0pt}{3.5ex} \textbf{Total (single telescope)} & \textbf{80} \\
\hline
\end{tabular}
\end{center}
\end{table}

Table~\ref{tab:cost}'s six itemised components sum to approximately 70\,M\texteuro\ per telescope (2026 prices), to which we add a 10\,M\texteuro\ (15\%) contingency, giving the adopted total of 80\,M\texteuro. As a cross-check, we also consider a more optimistic scenario in which further design optimisation and production of more than one unit reduce the core, pre-contingency cost to around 60\,M\texteuro; applying a larger, 30\% contingency margin -- reflecting the added uncertainty of this more aggressive cost-reduction assumption -- to this `cheap' telescope again yields approximately 80\,M\texteuro. The two estimates, a conservative itemised baseline and an optimistic higher-margin alternative, converge on the same order-of-magnitude figure, which we therefore adopt as our working per-telescope cost estimate. Scaled to four telescopes, this yields a telescope-only cost of order 320\,M\texteuro, before any project-level costs, infrastructure connection, or shared VLTI upgrades discussed in Sec.~\ref{sec:cost-scope}.

Comparing this figure to the historical benchmarks of Sec.~\ref{sec:cost-history}, the proposed concept is cheaper than a conventional $f/1.8$ 8\,m-class telescope but not dramatically so, reflecting the fact that mirror segmentation, adaptive correction, and the coud\'e relay train add cost even as the reduced telescope mass saves on the structure and enclosure. The comparison nonetheless indicates that a four-telescope VLTI expansion along these lines is cost-competitive with, and substantially smaller in absolute terms than, the capital expenditure of an ELT-class facility.

\section{CONCLUSION AND OUTLOOK}
\label{sec:conclusion}

We have presented an end-to-end cost and feasibility assessment for extending the VLTI with four additional 8\,m-class Unit Telescopes, an upgrade we refer to as the New Generation VLTI (ngVLTI). Building on the scientific case summarised in Sec.~\ref{sec:science}, we first showed that the resulting dense, homogeneous \textit{uv} coverage would turn the VLTI into a genuine imaging array: a worked reconstruction of a simulated young-stellar-object disk (Sec.~\ref{sec:facility-imaging}) recovers sub-au ring structures at milliarcsecond resolution directly from the visibilities, without parametric assumptions. Because each such image is reconstructed at a single wavelength, repeating the procedure across the many spectral channels of a dispersed instrument would deliver spectral-imaging cubes -- a capability entirely new to the VLTI, bringing it close to ALMA in its ability to map morphology and kinematics jointly, and doing so at near-infrared wavelengths and milliarcsecond resolution.

To deliver this array we described a compact telescope concept -- a fast ($f<1$), segmented, parabolic primary feeding a coud\'e relay compatible with the existing VLTI infrastructure -- and showed by ray-tracing that it achieves diffraction-limited performance across a 1\,arcmin field of view (Sec.~\ref{sec:optics}). We examined the mechanical trade-offs behind the adopted 60-segment, 1.2\,m primary and confirmed the structural feasibility of an ELT-heritage segment-support design at this scale (Sec.~\ref{sec:segmentation}), and we quantified the gravitational flexure of the telescope structure as it tracks in elevation (Sec.~\ref{sec:flexure}). The predicted focus, coma, and spherical-aberration terms reach hundreds of microns of wavefront error -- one to two orders of magnitude larger than on the stiffer $f/1.8$ UTs -- so the low-order component must be continuously offloaded from the adaptive-optics deformable mirror onto slower, larger-stroke actuators upstream.

Where that bulk correction is offloaded is not a neutral choice for an interferometer, and it emerged as one of the central technical tensions of this concept. Offloading onto the M2 hexapod is optically exact, since M2 is a single continuous mirror, but it requires physically moving the M2 mount, and any dynamical content in that motion couples directly into the optical path length relayed to the beam-combination laboratory, degrading fringe stability -- precisely the effect that led GPAO to discard its own fast M2 offload. Offloading instead onto the segmented M1 keeps the correction upstream of the coud\'e relay and leaves the M2 mount undisturbed, but a rigid-body segment motion can null only the local mean and slope of the target wavefront, not its curvature, leaving behind a periodic segment-scale residual known as \emph{scalloping}. A simple $\sim\!1/N$ scaling suggests this residual is at the percent level for our 60-segment array, so that M1 offload is well suited to the bulk, quasi-static flexure while the M2 hexapod, free of scalloping but noise-sensitive, remains the more exact actuator of last resort for the fastest corrections. The two channels are complementary rather than exclusive, following the same offload architecture originally intended for GPAO.

On the basis of this concept, an itemised, order-of-magnitude cost estimate places the per-telescope cost at approximately 80\,M\texteuro\ (2026 prices, including a first-of-a-kind contingency margin), or roughly 320\,M\texteuro\ for the four telescopes, before shared infrastructure and project-level costs (Sec.~\ref{sec:cost}). Benchmarked against Keck, Gemini, the VLT UTs, and the ELT, this figure is cheaper than a conventional 8\,m unit yet, as expected, not dramatically so, since segmentation, adaptive correction, and the coud\'e train recover part of the mass savings as complexity.

\vspace{0.5em}\noindent\textbf{Outlook.} Several elements of this assessment remain to be refined, and are the subject of ongoing work. The trade-off between segment size, areal fill-factor, and cost merits further optimisation, including alternative, non-hexagonal tiling schemes (Sec.~\ref{sec:segmentation}). Segment accessibility for maintenance -- e.g.\ whether M1 can be serviced with a man-lift from below, as on the existing UTs -- has not yet been assessed, nor has the error budget been finalised: with a primary roughly five times smaller than the ELT, it remains open whether individual error terms simply scale by that factor, and whether the resulting non-circular aperture and segment-gap diffraction pattern are acceptable for interferometric imaging. The M1-versus-M2 offload trade-off in particular calls for a quantitative evaluation of the scalloping residual against the specific flexure amplitudes of Table~\ref{tab:wfe}, coupled to a detailed M2-hexapod control model, in order to derive a post-correction residual wavefront-error budget and to confirm that the residual fringe perturbation is compatible with interferometric operation. The full optical path from the telescope down to the coud\'e room and the interface with the existing delay lines requires a complete design. A first side-by-side imaging comparison between the current 4-UT and proposed 8T arrays is given in Sec.~\ref{sec:facility-imaging}; extending this analysis to the multi-wavelength case is needed to quantify the fidelity of the spectral-imaging cubes that motivate the upgrade.

Taken together, these results indicate that a four-telescope expansion of the VLTI is both technically feasible and cost-competitive with historical facilities of comparable scale. By turning the VLTI into a true imaging -- and ultimately spectral-imaging -- array, such an upgrade would provide a long-term perspective for Paranal Observatory as a forefront facility for high-angular-resolution astronomy in the era of the ELT.

\section*{APPENDIX A. \MakeUppercase{UV-plane coverage of the proposed configuration}}
\label{sec:appendix-uvplane}

\begin{figure}[htp]
\begin{center}
\includegraphics[width=\linewidth]{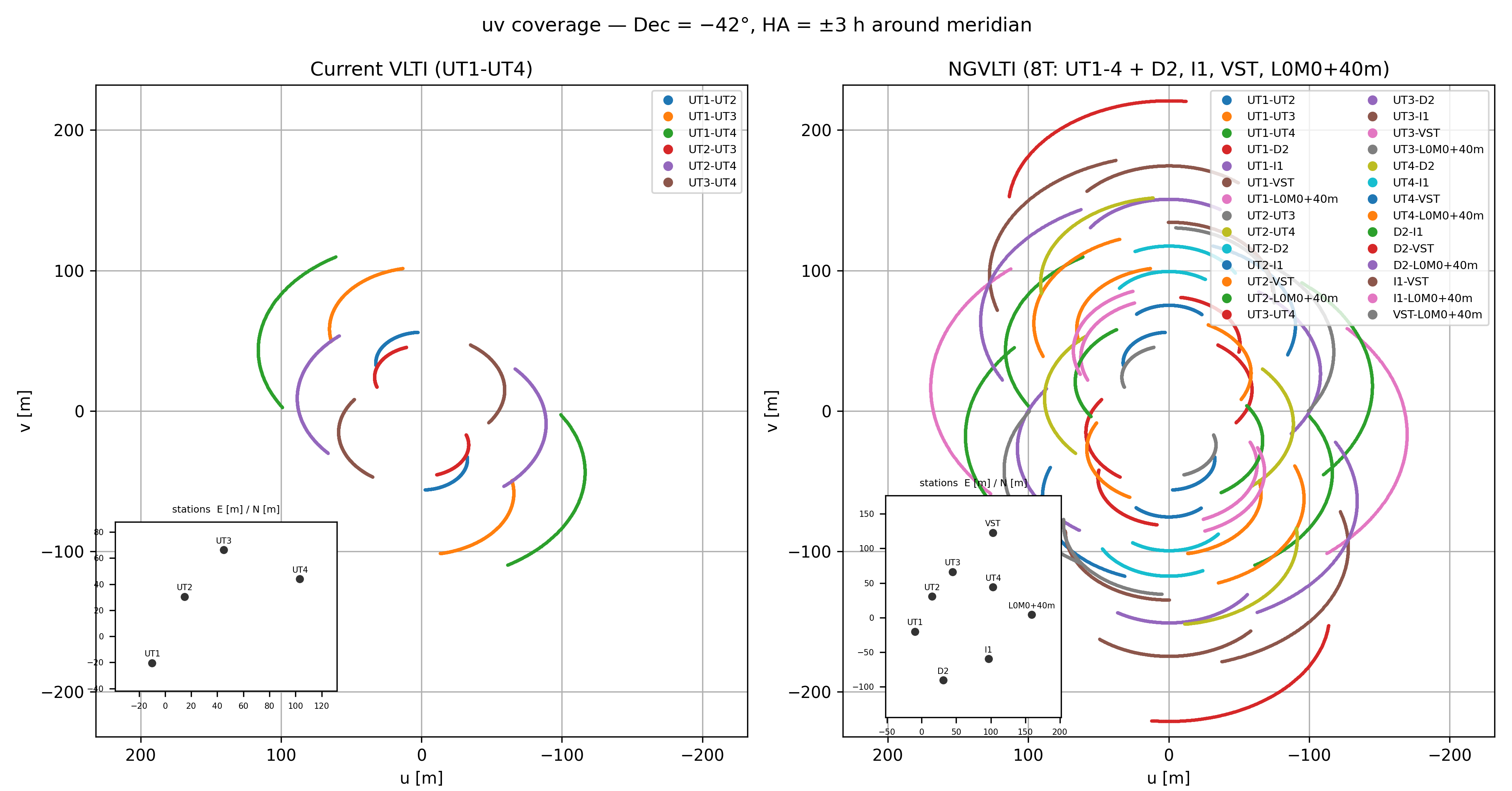}
\caption{\textit{uv}-plane coverage at declination $-42^\circ$ for $\mathrm{HA}=\pm3$\,h around meridian transit. \textit{Left:} the current VLTI array (UT1--UT4), giving 6 baselines. \textit{Right:} the proposed ngVLTI configuration, adding the D2 and I1 AT stations, the VST platform, and a station 40\,m beyond M0 along the L0$\to$M0 direction (Sec.~\ref{sec:facility-siting}), giving $\binom{8}{2}=28$ baselines. Insets show the ground layout of the stations in local East/North coordinates.}
\label{fig:uvtracks}
\end{center}
\end{figure}

Figure~\ref{fig:uvtracks} shows the \textit{uv}-plane coverage of the current and proposed VLTI arrays on its own, for the same source (declination $-42^\circ$) and hour-angle range ($\mathrm{HA}=\pm3$\,h around meridian transit) used in the imaging comparison of Sec.~\ref{sec:facility-imaging}. Each coloured track is the elliptical arc swept by one baseline in the $(u,v)$ plane as the Earth rotates, together with its conjugate ($-u,-v$); the inset in each panel shows the corresponding ground layout of the stations in local East/North coordinates.

\bibliographystyle{spiebib}
\bibliography{sample}

\end{document}